\documentclass[preprint,showpacs,showkeys,aps,prb]{revtex4}
\usepackage[dvips]{graphicx}

\begin{document}

\title{Linking Microscopic Reversibility to Macroscopic Irreversibility,
Emphasizing the Role of Deterministic Thermostats and \\
Simple Examples, At and Away from Equilibrium} 

\author{
Wm. G. Hoover and Carol G. Hoover               \\
Ruby Valley Research Institute                  \\
Highway Contract 60, Box 601                    \\
Ruby Valley, Nevada 89833                       \\
}

\date{\today}



\vspace{0.1cm}

\begin{abstract}
Molecular Dynamics and Statistical Mechanics make possible a particle-based
understanding of Thermodynamics and Hydrodynamics, including the fascinating 
Loschmidt contradiction between time-reversible atomistic mechanics and the
time-irreversible thermodynamic dissipation incorporated into macroscopic fluid
and solid mechanics.

\end{abstract}

\maketitle

\section{Introduction}

Among the many problem areas in mechanics the study of instabilities and irreversible
processes seem particularly interesting.  Engineering mechanics exists as a discipline
because failures of structures cost so many lives.  The analysis of local Lyapunov
instability gives a means for localizing and predicting catastrophic failures so that
there is a decidedly practical engineering  aspect of this fascinating scientific 
research area.

The goal we pursue here is to develop models that help us to understand.  It is profoundly interesting
that the small scale microscopic models of material behavior (ordinary Newtonian mechanics)
are time-reversible while the macroscopic  models of the same thing (finite-element and
finite-difference fluid mechanics and solid mechanics) are irreversible.  The tools from
nonlinear dynamics and chaos are useful in analyzing these two kinds of description.

Here we describe the basic building blocks for particle simulation and point out the ways
that these time-reversible simulations already lead to time-irreversible behavior.  Most of
the examples treated here are also described in our three books on computational statistical
mechanics, smooth-particle applied mechanics, and time reversibility, computer simulation,
algorithms, and chaos\cite{b1,b2,b3}.

\section{Algorithm for Conservative Particle Mechanics}

No special tricks are necessary to get started with particle-based simulations.
Microscopic mechanics can provide us with accurate particle trajectories $\{ \ q(t) \ \}$ .
All we need to do is to integrate Newton's ordinary differential equations of motion ,
$$
\{ \ m\ddot q = F(q) = -\nabla \Phi \ \} \ .
$$
Here $\Phi$ is the potential energy, a function of the coordinates $\{ \ q \ \}$ .  Alternatively,
we can obtain an equivalent coordinate-momentum description
$\{ \ q(t),p(t) \ \}$ by integrating Hamilton's first-order ordinary differential equations :
$$
\{ \ \dot q = +(\partial {\cal H}/\partial p) \ ; \
     \dot p = -(\partial {\cal H}/\partial q) \ \} \ .
$$
Both these approaches are time-reversible.  That is, a movie of the motion, played
backwards, satisfies exactly the {\it same} equations (with the values of $q$ and $p$ in reversed 
order and with the sign of $p$ changed also).  A movie is an excellent analog of numerical simulation.
Both the simulation and the movie are sets of discrete records of coordinates at discrete values
of the time, separated by the ``timestep'' $\Delta t$ .  In addition to the basic algorithm keep
in mind that three crucial questions remain to be answered: [1] what are the initial conditions, [2]
what are the boundary conditions, and (most important of all) [3] what is the problem to be solved?

Macroscopic continuum mechanics is based on the three conservation laws for mass, momentum, and energy :
$$
\dot \rho = -\rho\nabla \cdot u \ ; \
\rho \dot u = -\nabla \cdot P \ ; \
\rho \dot e = -\nabla u: P - \nabla \cdot Q \ .
$$
Here $\rho$ is density, $u$ is velocity, $e$ is energy per unit mass, $P$ is the pressure
tensor (force per unit area, necessarily a symmetric second-rank tensor), and $Q$ is the
heat flux vector (energy flow per unit area).  All these variables are continuous functions of
space and time.  Both $P$ and $Q$ , as well as $\dot \rho$ , $\dot u$ , and $\dot e$ , are
defined in the {\it comoving} frame, a coordinate frame
moving with the local velocity $u(r,t)$ .  Finite-difference approximations to the gradients
on the righthand sides of the three continuum equations, evaluated at a discrete set of spatial mesh
points, reduce the partial differential equations to ordinary ones, which can then be solved with
Runge-Kutta integration.  Again, the hard part of the problem is the same: what to do and how to
implement the initial and boundary conditions.

Different materials can be described by different types of constitutive relations (elastic,
plastic, viscous, ...) giving $P$ and $Q$ in terms of the basic $\{ \ \rho,u,e \ \}$ set,
together with their time derivatives and spatial gradients.  Time-reversed movies of solved
macroscopic problems look ``funny'' and make no sense.  This is because the underlying
phenomenological constitutive relations are typically irreversible.  The simplest most
familiar irreversible examples are Newtonian viscosity and Fourier heat conduction :  
$$
P = [ \ P_{\rm eq} - \lambda \nabla \cdot u \ ]I - \eta [ \ \nabla u + \nabla u^t \ ] \ ; \
Q = -\kappa \nabla T \ .
$$
In the symmetrized velocity gradient $\nabla u^t$ is the transpose of $\nabla u$ .  $I$ is the
unit tensor.  A boxed conducting fluid, with that fluid initially in motion, (a Rayleigh-B\'enard flow, for
instance, but with the box suddenly insulated and with the accelerating gravitational field
suddenly switched off) described with a shear viscosity $\eta$ and a heat conductivity $\kappa$
eventually comes to an isothermal state
of rest.  Evidently the reversed movie of this decay makes no sense and would correspond to an
illegal ``something from nothing'' contradicting the Second Law of Thermodynamics.  Of course,
with the right initial conditions and the right boundary conditions, one can indeed observe tornadoes!  For an
$L \times L$ (two-dimensional) system, with kinematic viscosity and thermal diffusivity of order $D$ the initial
gradients decay exponentially, $\simeq e^{-Dt/L^2}$ .  The reversed movie, with its exponential
growth, $\simeq e^{+Dt/L^2}$ , is simply wrong.

In applications of mechanics to simulation we strongly recommend the use of the fourth-order
Runge-Kutta integrator because it is easy to use and to modify for the treatment of open
systems interacting with their environments.  If the focus is on the time-reversibility of
conservative Newtonian systems it is useful to consider a very simple, yet rigorously time-reversible,
 integrator discovered by Levesque and Verlet\cite{b4}.  In order better to understand the coexistence of the
reversible microscopic and irreversible macroscopic views we adopt Levesque and Verlet's
``bit-reversible'' algorithm.  This approach generates a numerical trajectory in an {\it integer}
coordinate space, by rounding off the acceleration terms :
$$
\{ \ q_{n+1} - 2q_n + q_{n-1} = [ \ (F(q_n)(\Delta t)^2/m \ ]_{\rm integer} \ \} \ .
$$
The subscripts indicate the time, in units of the (integer) timestep $\Delta t$ .  The initial
conditions to start this algorithm are the coordinates at two successive times.

A simple illustration of the algorithm follows a harmonic oscillator trajectory, using unit mass, force constant,
and timestep $\Delta t$ :
$$
q_+ -2 q_0+ q_- = -q_0 \longrightarrow q_+ \equiv q_0 - q_- \ .
$$
The solution of repeating coordinates $\{ \ +1, +1, 0, -1, -1, 0, \dots \ \}$ \ , is typical, and
illustrates
the fact that no matter what the initial conditions, the solution is both periodic (for chaotic
problems, the length of
the period is of order the square root of the number of states) and reversible.  The algorithm
is a faithful analog of classical deterministic time-reversible mechanics.  If momenta are desired
they too can be approximated accurately from the coordinate values :
$$
p_0 \equiv \left[\frac{4}{3}\right]\frac{(q_+ - q_-)}{2\Delta t} - 
           \left[\frac{1}{3}\right]\frac{(q_{++} - q_{--})}{4\Delta t} \ .
$$
Figure 1 compares the energy calculated with these momenta with calculations based on the Beeman
and velocity-Verlet algorithms.  Our formulation (small dots in the Figure) is clearly an improvement.
It is evident that a promising research
area lies in the development of higher-order bit-reversible algorithms combining coordinates,
velocities, and accelerations from more than three successive times.

\begin{figure}
\vspace{1 cm}
\includegraphics[height=4cm,width=8cm,angle= 00]{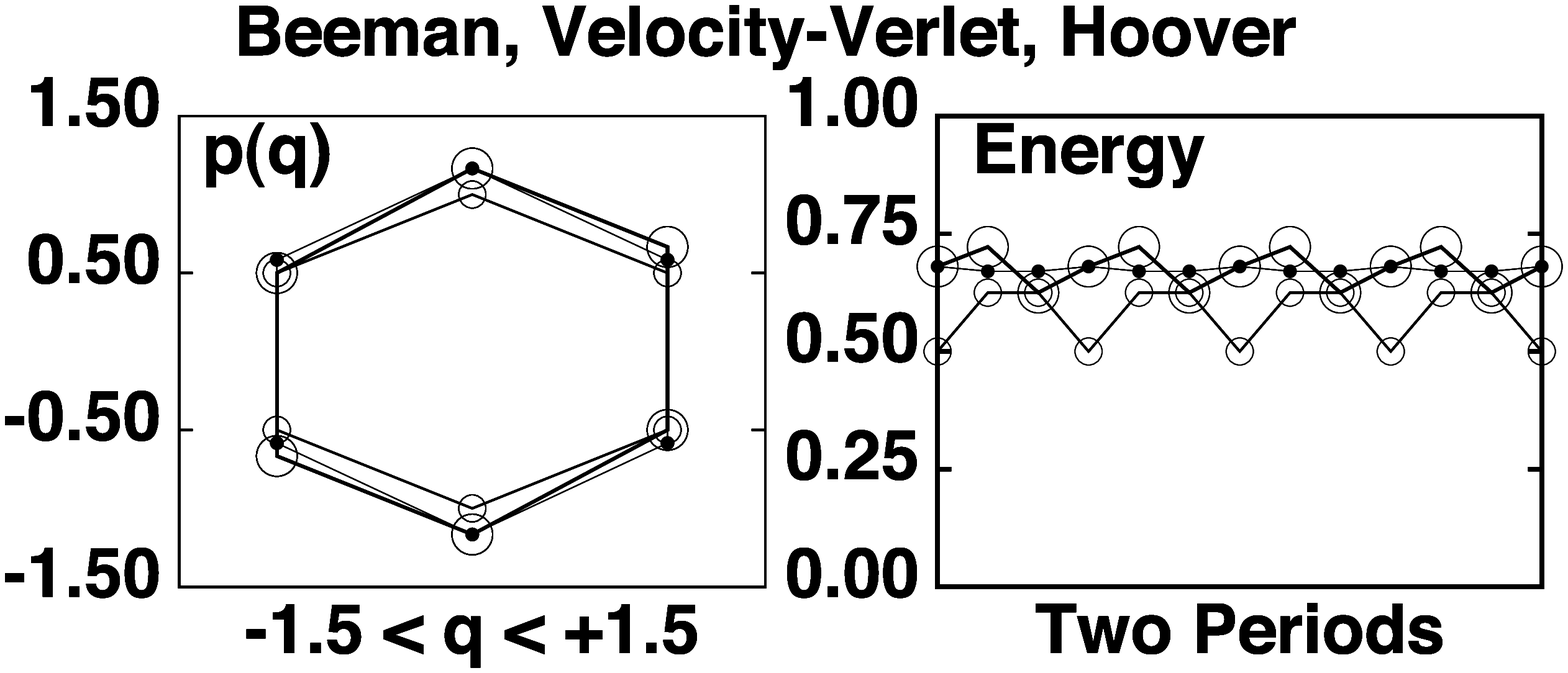}                   
\caption{ 
The Beeman and Velocity-Verlet approximations to the harmonic oscillator momentum are compared
with the more nearly accurate
formula given in the text (at the left).  The corresponding total energies are shown at the
right over two oscillations with $\Delta t=1$ .  The approximation mentioned in the text is the
best of the three and corresponds to the smallest dots.
}                                                                                                                    
\end{figure}
     
\section{Irreversibility from Time-Reversible Motion Equations?}

Is there any chance of detecting {\it irreversibility} with such a time-reversible algorithm?  Oddly enough, there
is!  It is based on the analysis of Lyapunov instability, looking in the {\it neighborhood} of the trajectory,
not just at the trajectory itself.  Such a nonlocal analysis necessarily depends upon the imagination
and the information contributed by an observer of the motion.  Let us take an example, a {\it maximally
irreversible} situation described by time-reversible, even bit-reversible,  Newtonian mechanics.  Consider the
pair of shockwaves
launched by the collision of two mirror-image fluid samples.  See Figure 2 for four snapshots of such a problem.
{\it Initially} the velocities are $\pm u$ .  {\it Eventually} the periodic $L \times L$ system shows
no more systematic motion -- the initial kinetic energy has been completely converted to internal energy (heat) :
$$
(u^2/2) \rightarrow e \ .
$$
Just as before, any portion of the developing trajectory can be reversed precisely and exactly despite the
Lyapunov instability (exponential growth of perturbations) of the dynamics.  

\begin{figure}    
\vspace{1 cm}
\includegraphics[height=20cm,width=10cm,angle= -90]{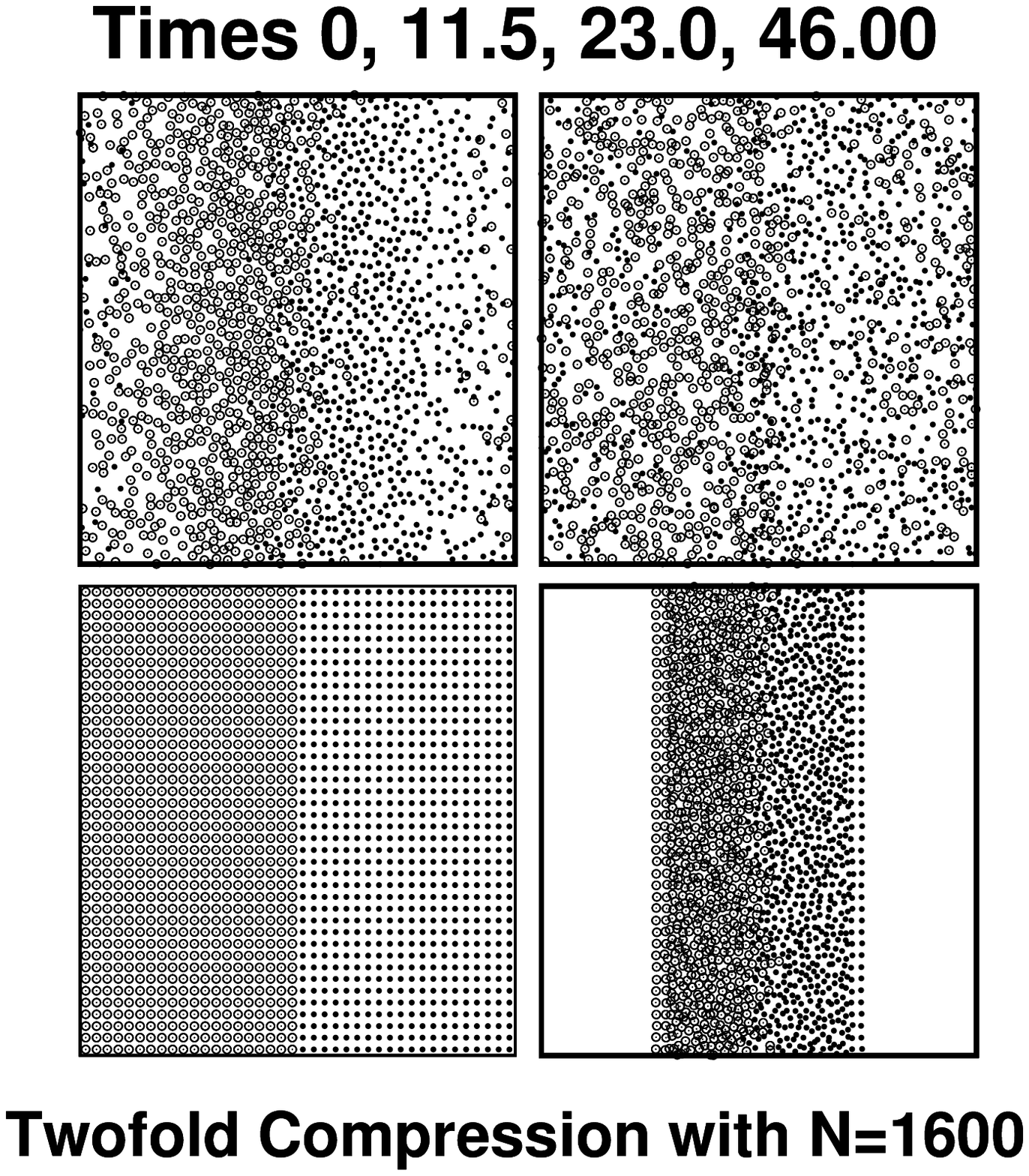}                                           
\caption{ 
Snapshots from the twofold compression of a $40 \times 40$ cold crystal with unit density and with
the pair potential $\phi = (1 - r^2)^4$ .  Initially half
the particles move to the right and half to the left, at speed 0.875 . The second snapshot, time =
11.5, and located at
bottom right, corresponds to the maximum twofold compression.  Such ``irreversible'' motions can be reversed
precisely, for as long as desired, with the Levesque-Verlet bit-reversible integration algorithm.
}
\end{figure}

There is a vast literature\cite{b5,b6,b7,b8,b9} on the quantification of Lyapunov instability, the exponentially
sensitive deformation of comoving hypervolumes in $q$ space, $p$ space, or $\{ \ q,p, \ \}$
phase space. For $N$ particles in two dimensions the $4N$-dimensional phase-space motion defines
$4N$ local Lyapunov exponents.  The fact that these ``local'' exponents
depend upon the chosen coordinate system can be viewed as a disadvantage or as an opportunity.
Again, there are many promising research problems suggested by this observation.  Optimizing the
analysis is certainly a useful and stimulating activity.

The largest Lyapunov exponent -- we will call it $\lambda_1(t)$ --  associated with the motion can be
found by following two nearby trajectories in time.  The primary or ``reference'' trajectory can be
generated with bit-reversible dynamics, so that it is possible to extend it as far as desired into
the future or the past\cite{b7}.  The dynamics of a nearby ``satellite'' trajectory is restricted by
constraining the satellite trajectory to stay within a fixed distance of the reference trajectory.  The
satellite dynamics can readily be generated with Runge-Kutta integration, rescaling the separation between
the two trajectories at the end of each time step.  The local Lyapunov exponent is :
$$
\lambda_1(t) \equiv \left[\frac{1}{\Delta t}\right] \ln 
\left[\frac{(\Delta r)_{\rm before}}{(\Delta r)_{\rm after}}\right] \ ; \
\Delta r \equiv |r_{\rm satellite} - r_{\rm reference}| \ .
$$
Although the motion equations are perfectly reversible for both the reference and the satellite, the
reversed satellite trajectory turns out to be totally unlike the forward one, {\it if} the system is a
nonequilibrium system.  Both the local Lyapunov exponent
associated with the instability and the identities of those particles making above-average contributions to
the offset vector separating the trajectories ,
$\Delta r \equiv r_{\rm satellite} - r_{\rm reference}$, are qualitatively different.  In a typical
shockwave simulation of the type shown in Figure 2 the
number of these more-heavily-weighted particles is about twice as great in the reversed motion as in the
forward one.  

This is an extremely interesting result.  No doubt it suggests various ``Arrows of Time'' which can be constructed
based on the structure of nearby trajectories (which react to the past, not the future).  A study
of irreversible flows from this standpoint should shed light on the reversibility paradox for
simple Newtonian and Hamiltonian systems.

\section{Irreversibility for Time-Reversible ``Open'' Systems}

In order to control nonequilibrium states, in particular nonequilibrium {\it steady} states, it is necessary
to do work and/or to exchange heat, with the system of interest.  The dynamics becomes a little more
complicated due to these interactions, but the interpretation compensates by becoming simpler.  Here we
take up the description of ``open systems''.

\subsection{Microscopic Pressure, Heat Flux, and Temperature}
``Open'' systems have mechanical and/or thermal connections to their environment, opening up the
possibility of simulating processes including thermodynamic work and the flow of heat.  Analysis of
these systems requires microscopic analogs for all the continuum variables.  Density, velocity, and
energy are the simplest of these.  We adopt the smooth-particle averaging method discovered by Lucy
and Monaghan in 1977\cite{b10,b11} :
$$
\rho(r,t) \equiv \sum_im_iw(r-r_i) \ ; \ \rho(r,t)u(r,t) \equiv\sum_im_iv_iw(r-r_i) \ ;
\ \rho(r,t)e(r,t) \equiv\sum_im_ie_iw(r-r_i) \ .
$$
The particle energies $\{ \ e_i \ \}$ are necessarily defined in the comoving frame, which moves
with velocity $u(r,t)$ .
A useful weight function, with a range $h$ which can be optimized, is Lucy's, here normalized for two
space dimensions :
$$
w(r) = (5/\pi h^2)(1+3z)(1-z)^3 \ z \equiv |r|/h \ .
$$
This weight-function approach guarantees the continuity of the first and second spatial derivatives
of the field variables and lends itself to optimization studies.

In addition to the basic mass, momentum, and
energy, several other variables need to be considered in order to compare microscopic and macroscopic
simulations.
Unlike their continuum cousins, the pressure tensor and the heat flux vector
from molecular dynamics are respectively even and odd functions of the time :
$$
PV = \sum_{\rm pairs}(rF)_{ij} + \sum_i (pp/m)_i \ ; \
QV = \sum_{\rm pairs}r_{ij}[ \ F_{ij}\cdot (p_i+p_j)/2 \ ] + \sum_i(ep/m)_i \ .
$$
Here $r_{ij} \equiv r_i - r_j$ and $F_{ij}$ is the force on Particle $i$ due to its interaction with
Particle $j$.  The individual particle energies $\{ \ e_i \ \}$ include half of each particle's pair
interactions with its neighbors.

Temperature needs a definition too.  The usual equilibrium definition, based on entropy, is useless
away from equilibrium where entropy has no consistent definition\cite{b1,b2}.  {\it At} equilibrium
{\it Temperature}
can be defined in many ways, all based on Gibbs' statistical mechanics or Maxwell and Boltzmann's
kinetic theory.  The even moments of the velocity distribution are examples.  In addition to these
there are also configurational definitions.  The simplest
``configurational temperature'' is based on an identity from Landau and Lifshitz' text\cite{b12,b13} :
$$
kT = \frac{\langle \ (\nabla {\cal H})^2 \ \rangle}{\langle \ \nabla ^2{\cal H} \ \rangle } \ .
$$
This definition follows from an integration by parts in Gibbs' canonical ensemble.  If the differentiation
indicated by $\nabla $ is carried out in momentum space the Landau-Lifshitz formula gives the
usual kinetic-theory definition of temperature , $mkT_{xx} = \langle \ p_x^2 \ \rangle$ .  If instead
the gradient is carried out in coordinate space the ``configurational temperature'' depends on
the first and second derivatives of the potential function governing the motion :
$$
kT_{\rm configurational} \equiv \langle \ F^2 \ \rangle / \langle \ \nabla^2 \Phi \ \rangle \ .
$$
One-body or many-body configurational temperatures, either scalar or tensor, can be defined in
this way.  But an evaluation of them for the shockwave problem reveals divergences.  Typically
particle values of $\nabla^2 \Phi$ frequently alternate between positive or negative values, so
that the corresponding configurational temperatures frequently diverge!  Configurational
temperature also has unphysical undesirable contributions arising from rotation whenever Coriolis' or
centrifugal forces are significant.

The simplest definition for temperature is the kinetic second-moment one.  It is based on a mechanical model of a
working ideal-gas thermometer.  In that instance a relatively heavy mass-$M$ ``system atom'' interacts
with a collection of light-weight mass-$m$ ``ideal-gas thermometer'' particles characterized by
an unchanging equilibrium Maxwell-Boltzmann distribution with temperature $T$ .  Kinetic theory shows
that the averaged
effect of such collisions causes the system-atom velocity to decay while its mean-squared velocity approaches
the equilibrium value for the temperature $T$ :
$$
\langle \ \dot v_x \ \rangle \propto -(v_x/\tau ) \ ; \
\langle \ \dot v_xv_x \ \rangle \propto [ \ (kT_{xx}/m) - v_x^2 \ ]/\tau \ ; 
 \ [ \ {\rm for} \ m<<M \ ] \ .
$$
Accordingly, we adopt the kinetic definition of temperature in what follows.  With temperature
defined we can proceed to devise ``thermostats'' able to control it.

\subsection{Time-Reversible Deterministic Thermostats}

The first of the deterministic mechanical thermostats was Woodcock's isokinetic thermostat\cite{b14},
implemented by rescaling the velocities at the end of each timestep.  Much later it was
discovered\cite{b15,b16} that a continuous time-reversible version of this thermostat could be
implemented with a {\it time-reversible} friction coefficient  $\zeta$ :
$$
\dot p = F(q) - \zeta p \ ; \ \zeta = \sum (F\cdot p)/\sum (p^2/m) \rightarrow (d/dt)\sum (p^2/m) \equiv 0 \ .
$$
This ``isokinetic'' thermostat can be applied to one or many particles and to one or many space directions.

An illustrative application is the ``Galton Board''\cite{b1,b2,b15}, in which a single particle is accelerated through a
lattice of scatterers but constrained to move at constant speed.  Overall, the potential energy drops.
Because the mean value of the friction
coefficient is necessarily positive, the phase-space probability density collapses onto a multifractal
strange attractor, quantifying the rarity of nonequilibrium phase-space states.  This approach to temperature control
is often termed the ``Gaussian'' thermostat because Gauss' Principle (of Least Constraint) gives this
thermostat when applied to the problem of constraining the kinetic energy\cite{b16}.  Reference 15 is
a detailed discussion of the model (summarized in References 1 and 2).  This work clearly shows the
fractal nature of the phase space (with vanishing phase volume) that results when
the dynamics is thermostated.  Fancier thermostats, based on statistical mechanics, can be found.

In 1984 Shuichi Nos\'e discovered a precursor of the best of them, a thermostat\cite{b17} with a
more elaborate basis
in Lagrangian and Hamiltonian mechanics, but somewhat disfigured by a novel
``time-scaling variable'' $s\equiv dt_{\rm old}/dt_{\rm new}$ .  His
thermostat imposed Gibbs' canonical phase space distribution at equilibrium rather than the less-usual
isokinetic  one.  A simplification of his equation of motion, without the useless time-scaling,
likewise contained a friction coefficient, which itself obeyed an evolution equation depending upon
{\it past} values of the kinetic energy :
$$
\{ \ \dot p = F(q) - \zeta p \ \} \ ; \ \dot \zeta = [ \ (T(\{ p \})/T_0) - 1 \ ]/\tau ^2 \ 
[ \ {\rm Nos\acute e-Hoover} \ ] \ .
$$
The relaxation time $\tau$ is a free parameter determining the time required for the thermostat forces
$\{ - \zeta p\}$ to bring the kinetic temperature $T(\{p\})$ to the desired thermostat temperature $T_0$ .
Just as in the isokinetic case the nonequilibrium averaged friction coefficient for this ``Nos\'e-Hoover''
mechanics is positive, leading once again to multifractal phase space distributions away from equilibrium.
There is an extensive somewhat mathematical literature having to do with picking the ``right''
relaxation time or the ``right'' thermostat.

The Gaussian and Nos\'e-Hoover thermostats are particularly useful for controlling nonequilibrium problems,
such as shear flows and heat flows, and the Rayleigh-B\'enard problem combining them.  The definition of
temperature depends upon the definition of the local velocity $u(r,t)$ .  A straightforward definition
of velocity, which nicely satisfies the continuity equation exactly\cite{b3}, can be based on
smooth-particle weighting functions :
$$
u(r,t) \equiv \sum_i v_iw(|r -r_i|)/\sum_iw(|r - r_i|) \ .
$$

The Gauss, Nos\'e, and Nos\'e-Hoover thermostats can {\it all} be related to Hamiltonian mechanics.
Dettmann, together with Morriss, carried out
much of this work, with later contributions by Bond, Laird, and Leimkuhler, and then by Campisi, H\"anggi,
Talkner, and Zhan\cite{b18,b19,b20,b21}.  All of them helped to clarify the connections of time-reversible thermostats with
standard Hamiltonian mechanics.
This work leads to the conclusion that {\it many} different thermostats can be used at equilibrium
but that some of them fail in nonequilibrium situations, even in situations close to equilibrium.
Just as in real life the failures, rather than the successes, are the more newsworthy subjects.  Let us
turn to some examples.

\subsection{Thermostat Failures  -- Oscillators, Heat Conduction, and the $\phi^4$ Model}

A very stimulating ``log-thermostat'' has just been described by Campisi, H\"anggi, Talkner, and
Zhan\cite{b21}.  They pointed out that the microcanonical (constant energy) ensemble distribution
for a logarithmic potential generates (at least formally) the Maxwell-Boltzmann velocity distribution :
$$
\phi \equiv kT\ln q \rightarrow \int_0^{+\infty}dq\delta[ \ 2H_0 - kT\ln q^2 - (p^2/m) \ ]
\propto \exp [ \ (H_0/kT) - (p^2/2mkT) \ ] \ 
$$
Because the dynamics of this thermostat is unstable, there being nothing to keep $q$ away from the
origin, in applications they recommend using $kT\ln(q^2 + \delta^2)$ , where $\delta$ is sufficiently
small.

Our effort to use this thermostat for a nonequilibrium heat flow problem failed.  Connecting a cold
and a hot log-thermostat to opposite ends of a two-particle $\phi^4$ chain gave different temperatures
at the two ends, but {\it no heat flux at all}.  The problem is that the Hamiltonian log-thermostat is unable
to replicate the phase-space contraction associated with dissipative systems.  There are some other
examples of such failures.  Leete and Hoover's Hamiltonian\cite{b22,b23} ,
$$
{\cal H}_{\rm HL} = \sqrt{4K(p)K_0} + \Phi(q) - K_0 \ ,
$$
keeps the kinetic energy, $\sum (m\dot q^2/2)$ constant, equal to $K_0$ .  The {\it configurational}
temperature can alternatively
be kept constant using a special Hamiltonian.  In both these cases a cold and a hot thermostated region,
in contact with Newtonian regions, gives {\it no heat flux at all} despite huge temperature differences.  The
lesson is that Hamiltonian mechanics is not able to describe dissipation properly.

\section{Thermostat Successes'' Oscillators and Complex Systems}

A ``good'' thermostat should, for instance, be able to provide good solutions of the Rayleigh-B\'enard problem,
heat transfer through a compressible fluid in a gravitational field.  It should also be useful in treating
small-scale ``toy problems''.  The simplest thermostat test problems
are [1] a harmonic oscillator\cite{b24} with a coordinate-dependent temperature :
$$
T(q) = 1 + \epsilon \tanh(q) \ ;
$$
and [2] the flow of heat through a $\phi^4$ chain of particles :
$$
{\cal H} \equiv \sum_i [ \ (p_i^2/2m) + (\delta x_i^4/4) \ ]  + \sum_{i<j} \phi_{ij} \ ,
$$
where $\phi$ is a nearest-neighbor Hooke's-Law potential and where the first few and last few particles
in the chain are thermostated with a Gaussian or Nos\'e-Hoover or another thermostat.  This model is a
specially good one to study because it is known to satisfy Fourier's law, even in one dimension.  A comparison
of seven different thermostat methods showed that the $\phi^4$ problem is well-posed and relatively easy to
solve\cite{b25}.

Some Hamiltonian-based thermostats are ineffective for nonequilibrium problems\cite{b21,b22,b23}
and it is useful to
understand why.  At equilibrium a given temperature and volume imply corresponding values of the
kinetic and potential energies.  This is also true for particular states away from equilibrium,
even where there is no longer a unique
equation-of-state relation.  Using a Hamiltonian thermostat away from equilibrium one can independently
specify the kinetic energy {\it and} the potential energy or the temperature {\it and} the energy.
This additional freedom contradicts the notion of thermodynamic state and can
lead to very strange results\cite{b23}.  Constraining the configurational temperature or using a version of
Hamiltonian mechanics to constrain the kinetic energy discovered by Hoover and Leete provide temperature
profiles that make no sense.  The log-thermostat is another demonstration that Hamiltonian mechanics
is a poor choice for thermostats.  This paradoxical situation is the symptom of two incompatible requirements on the
dynamics: [1] Liouville's Theorem requires that the phase-space motion be incompressible; [2] Heat
flow consistent with the Second Law of Thermodynamics requires that the phase volume decrease to zero.

Several of the thermostats have no problem with generating heat flows and solve the problem of
decreasing phase volume by generating strange attractors in the phase space.  Let us consider what
might appear to be the
simplest of these problems, the Nos\'e-Hoover oscillator\cite{b24} with a temperature gradient\cite{b1} :
$$
\dot q = p \ ; \ \dot p = -q - \zeta p \ ; \ \dot \zeta = p^2 - T(q) \ ; \
T(q) = 1 + \epsilon \tanh(q) \ ; \ 0 < \epsilon < 1 \ .
$$
For $\epsilon > 0.4$ the motion is a one-dimensional limit cycle with $\langle \ \zeta \ \rangle$
positive.  The mean value of the friction coefficient $\zeta$ in the range $(0.44 < \epsilon < 1.0)$
increases from about
0.15 to 1.35 . Figure 3 shows the gradual expansion of the hysteretic limit cycle as the maximum
temperature gradient
is increased from 0.44 to 1.00 .

Figures 4 and 5 show a bit of the complexity associated with smaller values of
the temperature gradient.  Using an initial momentum of unity gives  more regular attractors, of the type
shown to the left.  On the other hand, much higher initial momenta give chaotic distributions like
those shown to the right.  This complexity
is no doubt related to that seen without any temperature gradient at all\cite{b24}.  In that latter case the
phase-space distribution is divided into an {\it infinite} number of coexisting distributions, whose union
is Gibbs' canonical distribution ,
$$
f_{\rm Gibbs} = e^{-[ \ q^2 + p^2 + \zeta^2 \ ]/2} \ .
$$

The Figures show projections of a strange attractor that forms with $\epsilon = 0.40$ .
The Lyapunov spectrum in this case is nearly symmetric, so that it is difficult to compute
an accurate information dimension of the attractor.

\begin{figure}

\vspace{1 cm}
\includegraphics[height=8cm,width=4cm,angle= -90]{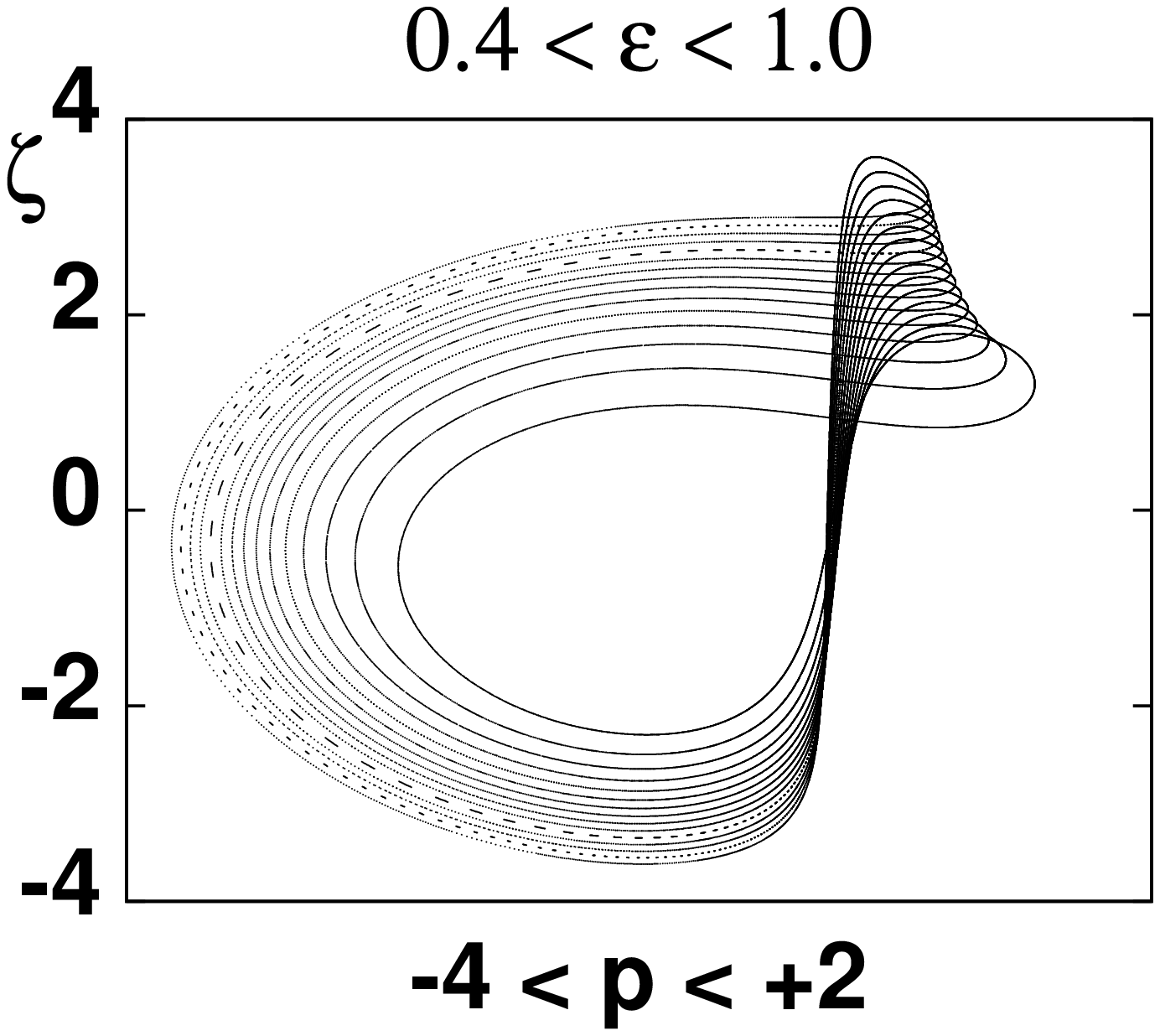}
\caption{
$\zeta(p)$ is plotted for fifteen equally-spaced values ($\epsilon =$ 0.44 to 1.00) of the maximum
temperature gradient for the $(q,p,\zeta)$ nonequilibrium Nos\'e-Hoover oscillator.  All these data
correspond to fully-converged limit cycles.
}
\end{figure}

\begin{figure}
\vspace{1 cm}
\includegraphics[height=12cm,width=6cm,angle= -90]{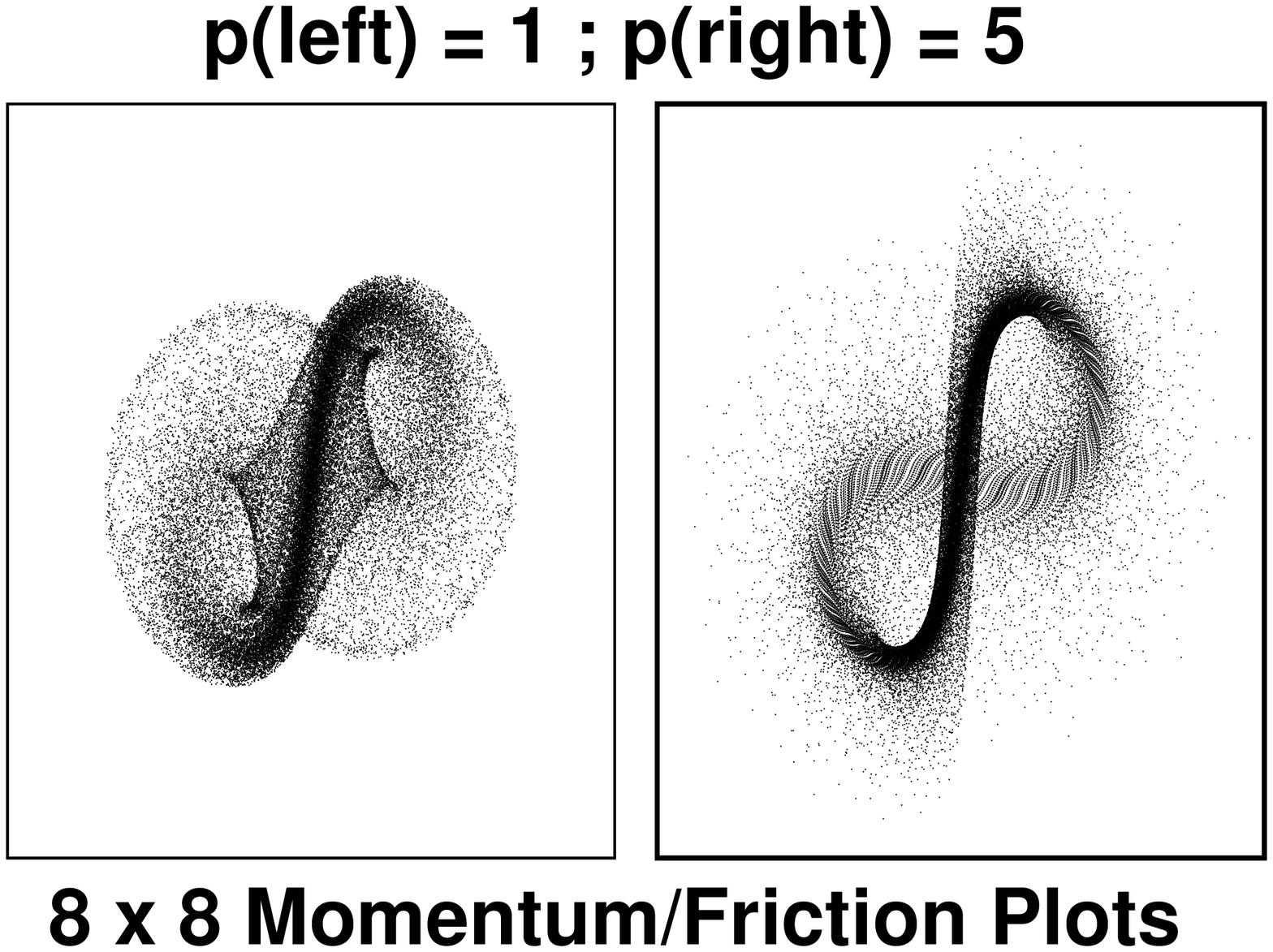}        
\caption{
$\zeta(p)$ for the nonequilibrium Nos\'e-Hoover oscillator $(\epsilon = 0.4)$ is plotted between
the limits $\pm4$ for two different initial conditions.  For $p=5$ the Lyapunov exponents are
roughly $\pm 0.0025$ and $0$.  For $p=1$ the exponents are much larger in magnitude, $\pm 0.0085$ and $0$.
}
\end{figure}

\vspace{1 cm}

\begin{figure}
\includegraphics[height=12cm,width=6cm,angle= -90]{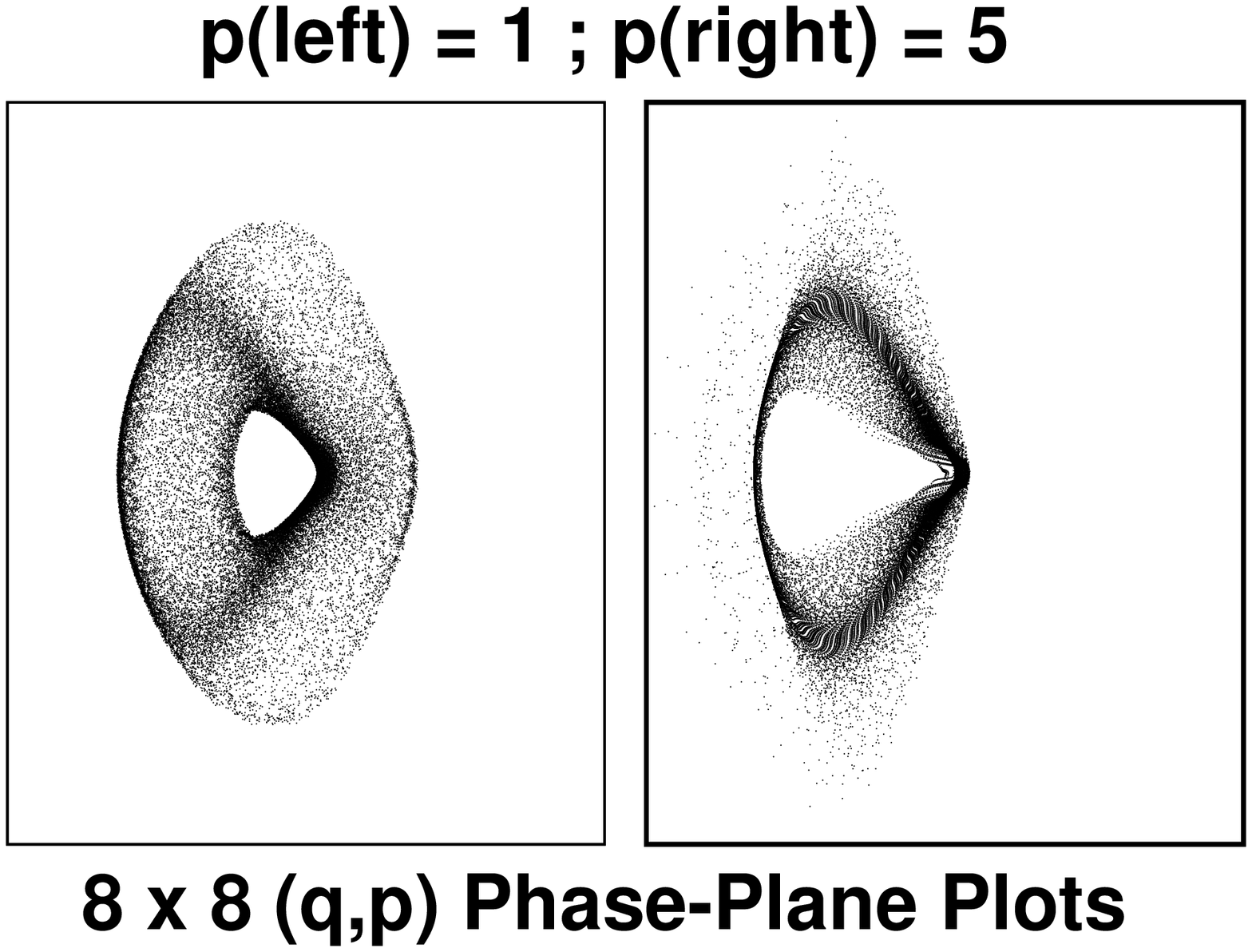}
\caption{
$p(q)$ is plotted between the limits $\pm4$ for the two different initial conditions of Figure 4 .
Note the preference of the oscillator for the lower-temperature states to the left of the origin. 
}
\end{figure}

Fortunately, the complex dynamics of the thermostated oscillator can be greatly simplified by
adding another control variable, a friction
coefficient controlling the fourth velocity moment\cite{b26} :
$$
\{ \ \dot q = p \ ; \ \dot p = -q - \zeta p - \xi p^3 \ ; \dot \zeta = p^2 - T(q) \ ; \
\dot \xi = p^4 - 3p^2T(q) \ ; \ T(q) = 1 + \epsilon \tanh(q) \ \} \ .
$$
At equilibrium the extra control variable allows the oscillator to sample the complete
canonical distribution.
This works at nonequilibrium too.  Figure 6 compares the distributions of the two friction
coefficients $(\zeta,\xi)$ for $\epsilon$ equal to 0.5 and 1.0 . Even in the latter case
the chaos induced by the two coefficients is enough to prevent collapse of the dynamics onto
a limit cycle.  Although counterintuitive, it appears to be true that a four-dimensional
attractor is actually much {\it simpler} than its three-dimensional counterpart.

\vspace{1 cm}

\begin{figure}                                                                       
\includegraphics[height=12cm,width=6cm,angle= -90]{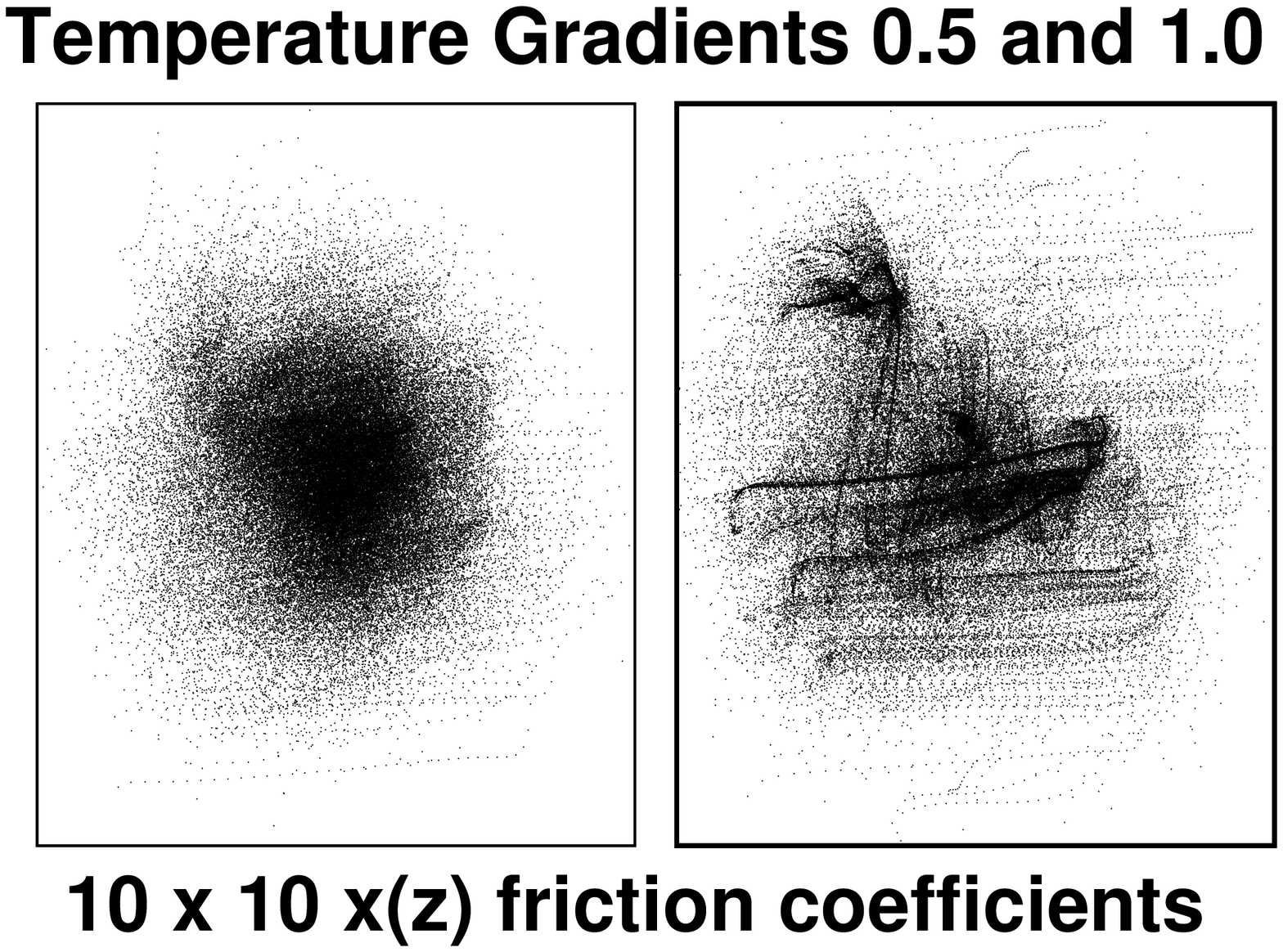}
\caption{ 
Friction coefficient distribution $\xi (\zeta )$ for two values of the maximum temperature
gradient, 0.5 and 1.0 .  This doubly-thermostated oscillator covers the complete canonical
distribution in the equilibrium case.  Here there is no dependence of the attractor on the
initial value of the momentum.  100,000 points are printed taken from the last half of a
200 million timestep run.  The timestep $\Delta t$ is 0.0002 .
}
\end{figure}

\subsection{Larger Systems and Thermodynamics}

Larger systems fit the pattern to which the small systems hint.  The phase-space distribution
shrinks to a strange attractor.  In a system with several thermostated degrees of freedom
Liouville's Theorem gives the details of the shrinkage\cite{b1,b15,b26} :
$$
(d\ln f/dt) \equiv -(\dot \otimes/\otimes) = \sum \zeta \equiv \exp[ \ (\dot S/k) \ ] \ .
$$
Here $\dot S$ is the external entropy production, the heat extracted from the controlled system
by the thermostats, divided by the thermostat temperature.  $\otimes$ is a small comoving
phase volume.  $\otimes$ has three possible evolutions: it can expand; it can shrink; or it can
remain the same.  The last possibility is the equilibrium one, with no net heat transfer to the
outside world.  The first possibility (expansion) is ruled out for steady states, as a continually
expanding phase volume implies catastrophic instability.  Only the possibility of continual shrinkage,
dissipation, is left.  The accessible phase-space states for a nonequilibrium steady state
continually decrease in number as the volume shrinks (exponentially fast) toward zero.  The
deterministic time-reversible thermostats make possible a simple geometric interpretation of
the Second Law of Thermodynamics.  Nonequilibrium steady states necessarily collapse to a
zero-volume strange attractor.  Thus nonequilibrium states are vanishingly rare.  Any attempt
to reverse the (time-reversible) dynamics would lead to divergence, with a positive Lyapunov sum,
and a violation of the Second Law.  What happens in fact is that, when reversed, the dynamics
soon breaks its time symmetry and seeks out again the attractor.  Time-reversible thermostats
have deepened our understanding of the Second Law\cite{b1,b27}.

\section{Summary}
The paradoxical reversibility properties of Newtonian and Hamiltonian mechanics can be modeled
with bit-reversible algorithms.  Such algorithms don't exist in cases where the phase volume
changes, where the mechanics is thermostated.  In the latter case Lyapunov instability seeks out
the unstable strange attractor, more stable still than is its repeller twin, leading to a simple
geometric understanding of the Second Law of Thermodynamics for open systems.

The symmetry breaking revealed by strong shockwaves suggests that a deepened understanding of
isolated systems can come from study of the local Lyapunov spectrum.  Both of these problem
areas, nonequilibrium conservative systems and nonequilibrium open systems, suggest
many interesting research opportunities for the future.

\section{acknowledgment}

This work represents a continuing effort which has been stimulated by  many colleagues.  Most
recently, we thank Francesco Ginelli and Massimo Cencini for inviting our contribution to a
special issue of the Journal of Physics A (Mathematical and Theoretical), “Lyapunov Analysis
from Dynamical Systems Theory to Applications”.  Anton Krivtsov and Vitaly Kuzkin encouraged
our participation in this meeting.  We are also specially grateful to Michele Campisi and his
colleagues and coworkers for many stimulating emails which helped us to identify some of the
many interesting problem areas presented here.

\end{document}